\documentclass{ifacconf}

\usepackage{graphicx}      
\usepackage{natbib}        

\usepackage{amsmath}
\usepackage{amssymb}

\usepackage{enumitem}

\newcommand{\refig}[2][Fig.]{#1~\ref{fig:#2}}   
\newcommand{\R}{{\mathbb{R}}}                   
\newcommand{\func}[2]{{#1}( {#2} )}             
\newcommand{\diff}{\mathrm{d}}                  
\newcommand{\od}[2]{
  \frac{\displaystyle \diff #1}{\displaystyle \diff #2}}   
\newcommand{\bm}[1]{\boldsymbol{#1}}            
\newcommand{\tp}[1]{#1^\top}                    
\newcommand{\z}{\bm{z}}
\newcommand{\cu}{\bm{u}}

\begin{document}
\begin{frontmatter}

\title{%
  State-Switching Control of\\
  the Second-Order Chained Form System
} 


\author[APU]{Masahide Ito} 

\address[APU]{%
  School~of~Information~Science~and~Technology,%
  ~Aichi~Prefectural~University,\\
  Aichi 480-1198, Japan 
  (e-mail: \texttt{masa-ito@ist.aichi-pu.ac.jp}).
}

\begin{abstract}                
This paper addresses a motion planning problem of the second-order chained form system.  The author presents a novel control approach based on switching a state.  The second-order chained form system is composed of three subsystems including two double integrators and a nonlinear system.  Switching a single state of the double integrators can modify the nature of the nonlinear system.  Such state-switching and sinusoidal control construct the basis of the proposed control approach.  The effectiveness is validated by a simulation result.
\end{abstract}


\end{frontmatter}

\section{Introduction}
\label{sec:intro}

Nonlinear dynamical systems with non-integrable differential constraints, the so-called nonholonomic systems, have been attracting many researchers and engineers for the last three decades.  A theorem in \cite{brockett_dgct83} gave a challenging and negative fact that there does not exist any smooth time-invariant feedback control law to be able to stabilize nonholonomic systems.  The applications include various types of robotic vehicles and manipulation.  Some of them have been often used as a kind of benchmark platform to demonstrate the performance of a proposed controller for not only a control problem of a single robotic system and also a distributed control problem of multiagent robotic systems.

A V/STOL aircraft without gravity~(\cite{hauser_autom92}), an underactuated manipulator~(\cite{arai_ieee-tra98}), and an underactuated hovercraft~(\cite{he_robotica16}) belong to a class of dynamic nonholonomic systems which are subject to acceleration constraints.  The mathematical representation of these systems can be transformed to the second-order chained form by a coordinate and input transformation.  The second-order chained form is a canonical form for dynamic nonholonomic systems.

Several control approaches to the second-order chained-form system have been developed so far.  Most of them focuses on avoiding the theorem of \cite{brockett_dgct83}.  \cite{ge_ijc01} and \cite{he_robotica16} exploit discontinuity in their stabilizing controllers; \cite{deluca_ijrr02} and \cite{aneke_ijrnc03} reduce the control problem into a trajectory tracking problem.  Other than those, \cite{yoshikawa_icra00} and \cite{ito_electron19} consider a motion planning problem (in other words, a feedforward control problem).

For motion planning of the second-order chained form system, this paper presents a novel control approach based on switching a state.
The second-order chained form system is divided into three subsystems.
Two of them are the so-called double integrators; the other subsystem is a nonlinear system depending on one of the double integrators.
In other words, the input matrix of the latter subsystem depends on a single state of the double integrators.
The double integrator is linearly controllable, which enables to switch the value of the position state in order to modify the nature of the nonlinear subsystem.  Steering the value into one corresponds to modifying the nonlinear subsystem into the double integrator; steering the value into zero corresponds to modifying the nonlinear subsystem into a linear autonomous system.  This nature is the basis of the proposed control approach.
The proposed approach is composed of such state-switching and also sinusoidal control inputs.  Its effectiveness is validated by a simulation result.

\section{Subsystem Decomposition of the Second-Order Chained Form System}
\label{sec:sub-syst-decomp_socf-sys}

Consider the following second-order chained form system:
\begin{equation}
  \label{eq:2nd-cf}
  \od{{}^2}{t^2}\bm{\xi} =
  \begin{bmatrix}
    1 & 0\\
    0 & 1\\
    \xi_2 & 0
  \end{bmatrix}
  \cu,
\end{equation}
where $\bm{\xi} := \tp{[\, \xi_1,\, \xi_2,\, \xi_3 \,]}$ and
$\cu = \tp{[\, u_1,\, u_2 \,]}$.
By defining a state vector as
$\z = \tp{[\, z_1,\, \dotsc, z_6 \,]} :=
\tp{[\, \tp{\bm{\xi}},\, \tp{\dot{\bm{\xi}}} \,]}$,
the system~\eqref{eq:2nd-cf} is described as
\begin{equation}
  \label{eq:affine-sys}
  \od{}{t} \z =
  \begin{bmatrix}
    z_4\\ z_5\\ z_6\\ 0\\ 0\\ 0
  \end{bmatrix}
  + \begin{bmatrix}
    0\\ 0\\ 0\\ 1\\ 0\\ z_2
  \end{bmatrix} u_1
  + \begin{bmatrix}
    0\\ 0\\ 0\\ 0\\ 1\\ 0
  \end{bmatrix} u_2.
\end{equation}
This affine nonlinear system~\eqref{eq:affine-sys} has equilibrium points
at $\z_e := \tp{[\, z_1^\star, \, z_2^\star, \, z_3^\star, \:
  0, \, 0, \, 0 \,]}$,\ $z_1^\star, z_2^\star, z_3^\star \in \R$
with $u_1 = u_2 = 0$.
By using the theorem of \cite{sussmann_siam-jco87},
we can easily confirm that the system~\eqref{eq:affine-sys}
(or \eqref{eq:2nd-cf}) is small-time local controllable at $\z_e$.

From the viewpoint to separate the control inputs,
the system~\eqref{eq:2nd-cf} is divided into the following two subsystems:
\begin{subequations}
\label{eq:subsysts}
\begin{align}
  \label{eq:1st-subsyst}
  \od{}{t}
  \begin{bmatrix}
    z_2 \\ z_5
  \end{bmatrix}
  &=
  \begin{bmatrix}
    0 & 1\\
    0 & 0
  \end{bmatrix}
  \begin{bmatrix}
    z_2 \\ z_5
  \end{bmatrix}
  +
  \begin{bmatrix}
    0 \\ 1
  \end{bmatrix}
  u_2,
  \\
  \label{eq:2nd-subsyst}
  \od{}{t}
  \begin{bmatrix}
    z_1 \\ z_3 \\ z_4 \\ z_6
  \end{bmatrix}
  &=
  \begin{bmatrix}
    0 & 0 & 1 & 0\\
    0 & 0 & 0 & 1\\
    0 & 0 & 0 & 0\\
    0 & 0 & 0 & 0
  \end{bmatrix}
  \begin{bmatrix}
    z_1 \\ z_3 \\ z_4 \\ z_6
  \end{bmatrix}
  +
  \begin{bmatrix}
    0 \\ 0 \\ 1 \\ z_2
  \end{bmatrix}
  u_1.
\end{align}
\end{subequations}
The subsystem~\eqref{eq:1st-subsyst} is
a second-order linear system---the so-called double integrator;
the subsystem~\eqref{eq:2nd-subsyst} looks a four-order linear system
but its input matrix~$\bm{b}_2$ depends on a state~$z_2$.
The subsystem~\eqref{eq:2nd-subsyst} is furthermore decomposed to the following subsystems:
\begin{subequations}
\label{eq:sub2systs}
\begin{align}
  \label{eq:subsyst11}
  \od{}{t}
  \begin{bmatrix}
    z_1 \\ z_4
  \end{bmatrix}
  &=
  \begin{bmatrix}
    0 & 1\\
    0 & 0
  \end{bmatrix}
  \begin{bmatrix}
    z_1 \\ z_4
  \end{bmatrix}
  +
  \begin{bmatrix}
    0 \\ 1
  \end{bmatrix}
  u_1,
  \\
  \label{eq:subsyst31}
  \od{}{t}
  \begin{bmatrix}
    z_3 \\ z_6
  \end{bmatrix}
  &=
  \begin{bmatrix}
    0 & 1\\
    0 & 0
  \end{bmatrix}
  \begin{bmatrix}
    z_3 \\ z_6
  \end{bmatrix}
  +
  \begin{bmatrix}
    0 \\ z_2
  \end{bmatrix}
  u_1.
\end{align}
\end{subequations}

The next section introduces the proposed control approach based on
the above-mentioned subsystem decomposition.

\section{State-Switching Control Approach}
\label{sec:stsw-contr-approach}

The section addresses a motion planning problem between equilibrium points of the second-order chained form system~\eqref{eq:2nd-cf}.

The author presents a control approach based on the subsystem decomposition that divides the second-order chained form system~\eqref{eq:2nd-cf} into the three subsystems~\eqref{eq:1st-subsyst}, \eqref{eq:subsyst11}, and \eqref{eq:subsyst31}.
The input matrix of the subsystem~\eqref{eq:subsyst31} depends on a state of the subsystem~\eqref{eq:1st-subsyst}, $z_2$.
The state $z_2$ can be constant by a control input $u_2$ because the subsystem~\eqref{eq:1st-subsyst} is linear controllable.

Now let us consider that $z_2$ between $0$ and $1$ is switched.
When $z_2$ is equivalent to $1$, the subsystems~\eqref{eq:subsyst11} and \eqref{eq:subsyst31} are completely same under a shared control input~$u_1$.
The shared control input brings that the state differences between those controlled subsystems then depend on their values at the moment when $z_2$ just becomes $1$
In meanwhile, when $z_2$ is equivalent to $0$, the subsystem~\eqref{eq:subsyst11} can be controlled independently of another subsystem~\eqref{eq:subsyst31}.

From this perspective, the author conceived the following basic maneuver for the above-mentioned motion planning problem:
\begin{enumerate}[
  label=\texttt{Step \arabic{enumi}}),%
  leftmargin=5em,labelsep=.5em,itemsep=0em%
]
\item 
  Steer $z_2$ from any initial value to $1$ by using $\func{u_1}{t} = 0,\ \func{u_2}{t} = \func{q_1}{t}$;
\item 
  Steer $z_3$ from any initial value to any desired value (in conjunction with it, $z_1$ is also driven) by using $\func{u_1}{t} = \func{q_2}{t},\ \func{u_2}{t} = 0$;
\item 
  Steer $z_2$ from $1$ to $0$ by using $\func{u_1}{t} = 0,\ \func{u_2}{t} = \func{q_3}{t}$;
\item 
  Steer $z_1$ from a value to any desired value by using $\func{u_1}{t} = \func{q_4}{t},\ \func{u_2}{t} = 0$;
\item 
  Steer $z_2$ from $0$ to any desired value by using $u_1 = 0,\ u_2 = q_5(t)$,
\end{enumerate}
where $q_i(t)$ is an appropriate sinusoidal function in the $i$-th phase ($i = 1, 2, \dotsc, 5$).
Note that sinusoidal steering is inspired by the result of \cite{ito_electron19}.

\begin{rem}
Some conventional approaches such as in \cite{aneke_ijrnc03} and \cite{hably_cdc14} also exploit similar (but different) subsystem decomposition.  Note that those approaches focus on not motion planning but stabilizing (via trajectory tracking).
\end{rem}

\section{Numerical Example}
\label{sec:num-ex}

A numerical example is shown to demonstrate the effectiveness of the proposed control approach.


Consider a motion planning problem between $\func{\bm{z}}{0} = \tp{[\, 3,\, 0.5,\, 1,\; 0,\, 0,\, 0 \,]}$ and $\func{\bm{z}}{4T} = \tp{[\, 0,\, 0,\, 0,\; 0,\, 0,\, 0 \,]}$.
The basic maneuver is adjusted as follows:
\begin{enumerate}[
  label=\texttt{Step \arabic{enumi}}),%
  leftmargin=5em,labelsep=.5em,itemsep=0em%
]
\item 
  Steer $z_2$ using from $0.5$ to $1$;
\item 
  Steer $z_3$ from $1$ to $0$;
\item 
  Steer $z_2$ from $1$ to $0$;
\item 
  Steer $z_1$ from a resultant value in \texttt{Step 2} to $0$.
\end{enumerate}
To execute this procedure, the following control inputs were adopted:
\begin{equation}
  \label{eq:contr-input4case2b}
  \left\{%
    \begin{array}{@{\,}ll@{\,}}
      \func{u_1}{t} = 0 \;\;\mbox{and} &\\[.25em]
      \multicolumn{1}{r}{%
      \qquad\func{u_2}{t} = \phantom{-}a_1 \omega^2 \sin{\omega t}},
      \quad &\mbox{for}\;\; t \in [0, T],
      \\[.75em]
      \func{u_1}{t} = -a_2 \omega^2 \sin{\omega t} &\\[.25em]
      \multicolumn{1}{r}{%
      \mbox{and}\quad \func{u_2}{t} = 0},
      \quad &\mbox{for}\;\; t \in (T, 2T],
      \\[.75em]
      \func{u_1}{t} = 0 \;\;\mbox{and} &\\[.25em]
      \multicolumn{1}{r}{%
      \func{u_2}{t} = -a_3 \omega^2 \sin{\omega t}},
      \quad &\mbox{for}\;\; t \in (2T, 3T],
      \\[.75em]
      \func{u_1}{t} = -a_4 \omega^2 \sin{\omega t} &\\[.25em]
      \multicolumn{1}{r}{%
      \mbox{and}\quad \func{u_2}{t} = 0},
      \quad &\mbox{for}\;\; t \in (3T, 4T].
    \end{array}
  \right.
\end{equation}

\refig[Figure]{case2b_time-plots} depicts 
a simulation result with $\omega = 2\pi$, $T = 1$, $a_1 = 0.5/(2\pi)$, $a_2 = a_3 = 1/(2\pi)$, and $a_4 = 2/(2\pi)$.
It is obvious that the desired motion is successfully planned.  The proposed control approach, therefore, was confirmed to be useful for motion planning.
The results of the conventional approaches such as in \cite{aneke_ijrnc03} and \cite{hably_cdc14} imply that the proposed approach can be (partially) combined with stabilization and trajectory tracking.
\begin{figure}[b]
  \centering
    \includegraphics[width=.5\textwidth]{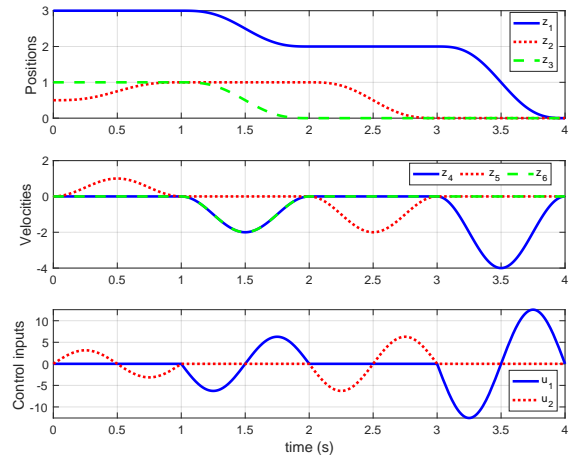}
  \caption{Time plots.} 
  \label{fig:case2b_time-plots}
\end{figure}
\section{Conclusion}
\label{sec:conc}

For a motion planning problem of the second-order chained form system, this paper has proposed a state-switching control approach based on subsystem decomposition.
The subsystem decomposition divides the second-order chained form system to three subsystems.
One of the subsystems has the input matrix that depends on a state of the other subsystem.
Switching the state between one and zero modifies the nature of the associated subsystem.
This is the key point of the proposed control approach.
The effectiveness of the proposed approach was shown by the simulation result.

Future directions of this study are:
\begin{itemize}
\item
  to compare the proposed approach with the other related ones;
\item
  to investigate further properties of the proposed approach; and
\item
  to extend the second-order chained form system into
  the higher-order one.
\end{itemize}


\bibliography{root}             
                                                   
\end{document}